\documentstyle[prb,aps,epsfig,multicol]{revtex}
\begin{document}
\topmargin 0.1cm
\sloppy
\draft
\title
{Specific heat of Bosons in a lattice}
\author{R. Ramakumar$^{a,}$\cite{add2} and A. N. Das$^{a}$}
\address{$^{a}$Condensed Matter Physics Group, 
Saha Institute of Nuclear Physics,
1/AF Bidhannagar, Kolkata-700064, INDIA \\
$^{b}$Department of Physics and Astrophysics, University of Delhi,
Delhi-110007, INDIA} 
\date{17 August 2005}
\maketitle
\begin{abstract}
  We present a theoretical study of 
specific heat of bosons ($C_v$) in  
a simple cubic lattice. We have studied the non-interacting bosons
and the Tonks gas.
For both cases, the $C_v$ above the bose 
condensation temperature shows considerable 
temperature dependence compared to that of free bosons.
For Tonks gas, we find that the low-temperature specific heat 
increases as the system gets closer to the Mott transition.
\end{abstract}
\begin{multicols}{2}
\narrowtext
\section{Introduction}
\label{sec1}
Investigations of collective properties of bosons in optical lattices
produced by counter propagating laser beams is now a rapidly expanding
field of research. It was shown\cite{jak} early on that bosons in 
optical lattices can be adequately modeled by employing a Bose-Hubbard 
model. Properties of a system modeled by this model depends 
on lattice symmetry, inter-site boson hopping energy ($t$), on-site 
boson-boson interaction strength ($U$), and the number of bosons per 
site ($n$). Experimentalists have achieved great 
control over several of these parameters by changing the characteristic
of the laser beams and their relative orientations. Recently, Greiner
and collaborators\cite{greiner} demonstrated bose-condensed to Mott 
insulator transition in a system of interacting bosons in a simple 
cubic optical lattice.
This transition was predicted in theoretical 
studies\cite{jak,fisher,nandini,shesh,monien,oost}
 on Bose-Hubbard model.
\par
For a system of repulsive interacting bosons living in a lattice,
bosons will try to avoid multiple occupancy of the sites. 
When the number of bosons per site is less than or equal to unity 
and in the strong interaction limit ($U>>t$) double (or multiple)
occupancy of any site is forbidden.
In this case, one has a quantum version of Tonks gas\cite{tonks} 
(here after referred to as Tonks gas) in a lattice. 
One dimensional  Tonks gas was investigated
by Girardeau\cite{gir} long time back.
Quite recently Tonks-Girardeau gas\cite{gir,lib} regime has been achieved in
experimental studies\cite{paredes}
of interacting bosons in one-dimensional optical
lattices. It is reasonable to expect that Tonks gas in two and
three dimensional lattice will be  achieved in the near future.
Thermodynamic properties of bosons in a lattice is of
considerable interest in this context.
In this short communication 
we present a theoretical study
of specific heat of non-interacting and
strongly interacting bosons in a simple cubic (sc) lattice.
As should be clear by now, the experiments mentioned above
and experiments that would be conducted in future are
our primary motivation for the work presented in this paper.
Heat capacity measurements on an optically trapped, strongly interactiong 
Fermi gas of atoms has been reported recently \cite{kinast} and 
one can expect that such measurements for bosons in an optical lattice 
will be done in the near future.
The effects of the presence of the harmonic confining potential is not 
included in our calculations given below. Our results 
for the non-interacting bosons would be reasonable for a trap of
shallow confining potential.
It may be mentioned, however, that the strongly interacting regime 
(Tonks limit) could be reached in
experiments with extremely deep optical lattice potential. This, in turn,
creates a deep harmonic trap, generated by the envelope of the laser beams. 
For realistic calculations of experimentally measured quantities in
the Tonks limit at present, one should take account of the harmonic trap 
potential. Our study in the Tonks limit may be considered as a 
starting point of a more complete calculation where the effects of
inhomogeneties produced by the confining potenetial is included.   
\section{Specific heat of lattice Bosons}
\label{sect2}
The specific
heat can be obtained from the energy given by
\begin{equation}
E=\int_{-W}^{W}\epsilon\rho(\epsilon)n_{B}(\epsilon) d\epsilon\,,
\end{equation}
where $\rho(\epsilon)$ is the normalized single boson Density of States (DOS)
of  a band of width $2W$, $n_{B}(\epsilon)=
1/[exp(\beta(\epsilon-\mu)-1]$ with  
$\beta=1/k_{B}T$ where $k_{B}$ is the Boltzmann's constant and $T$
is the temperature, and $\mu$ is the chemical potential. 
\par
First we consider the case of $T>T_{B}$, where $T_{B}$ is the
Bose condensation temperature.
In this case, the chemical potential is below the band bottom and
is temperature dependent. The chemical potential and the bose
condensation temperature are obtained by
solving the number equation
\begin{equation}
n = \int_{-W}^{W}
    \rho(\epsilon)n_{B}(\epsilon)d\epsilon ,
\end{equation}
where $n$ is the number of bosons per site.
For  $T>T_{B}$, the energy is given
by Eq. (1) and we obtain the constant volume specific heat
(per site) to be
\begin{equation}
C_{v}=\frac{1}{k_{B}T^{2}}(I_{3}-\frac{I_{2}I_{1}}{I_{0}}),
\end{equation}
where,
\begin{eqnarray}
I_{0}&=&\int_{-W}^{W}
 F(\epsilon-\mu)\rho(\epsilon)d\epsilon , \\
I_{1}&=&\int_{-W}^{W}
\epsilon F(\epsilon-\mu)\rho(\epsilon)d\epsilon , \\
I_{2}&=&\int_{-W}^{W}
(\epsilon-\mu)F(\epsilon-\mu)\rho(\epsilon)d\epsilon ,\\
I_{3}&=&\int_{-W}^{W}
\epsilon (\epsilon-\mu) F(\epsilon-\mu) \rho(\epsilon) d\epsilon . 
\end{eqnarray}
In the above equations,
\begin{equation}
F(\epsilon-\mu) = \frac{e^{\beta(\epsilon-\mu)}}
{\left[e^{\beta(\epsilon-\mu)}-1\right]^{2}}\,.
\end{equation}
\par
Now we consider the case of $T<T_{B}$. 
In this case the chemical potential is temperature independent and
is pinned to the bottom of the band ($-W$).
The band bottom
has a macroscopic number of particles (the Bose condensate)
and
the rest of the particles are in the excited states. Denoting
$n_{ext}$ as the number of bosons (per site) in excited states, the
energy per site is
\begin{equation}
E=\int_{-W}^{W}\epsilon\rho(\epsilon)n_{B}(\epsilon) d\epsilon
-W(n-n_{ext}).
\end{equation}
The specific heat in this case becomes,
\begin{equation}
C_{v} =\frac{1}{k_{B}T^{2}}\int_{-W}^{W}
(\epsilon+W)^{2}\rho(\epsilon)F(\epsilon+W)d\epsilon.
\end{equation}
\par
{\em Non-interacting bosons}: The Hamiltonian of 
the non-interacting bosons in a tight-binding
energy band is:
\begin{equation}
H=\sum_{{\bf k}}[\epsilon({\bf k})-\mu]c^{\dag}_{{\bf k}}c_{{\bf k}}\,,
\end{equation}
where,  
$\epsilon_{sc}(k_{x},k_{y},k_{z})=-2t\sum_{\nu=x}^{z}cos(k_{\nu})$
is the energy band structure
of a single boson in a sc lattice
with lattice constant set to unity,
$c_{{\bf k}}=\sqrt{1/N_{S}}\sum_{i}c_{i}\,exp(i\bf{k}.\bf{R_i})$,
$c_{i}$ is the annihilation operator of a boson at site $i$,
and $N_{S}$ is the number of lattice sites.
\par
{\em Strongly interacting bosons (the Tonks gas limit)}:
To study the Tonks gas limit, we start from the Bose-Hubbard model
given by
\begin{equation}
H= -t\sum_{<ij>}(c^{\dag}_{i}c_{j}+h.c.)
+\frac{U}{2}\sum_{i}n_{i}(n_{i}-1)-\mu\sum_{i}n_{i}\,,
\end{equation}
where $n_{i}=c^{\dag}_{i}c_{i}$.
In the above Hamiltonian
$c^{\dag}_{\bf i}$ ($c_i$) is the boson creation (annihilation)
operator,
$U$ the boson-boson repulsive interaction energy.
The Tonks gas regime
is achieved in the strong correlation limit ($U/t >> 1$) when 
the number of particles per site $n=N/N_{s}$ is restricted 
to be less than or equal to unity. In this limit, the
double (or multiple) occupancies of the sites are forbidden. 
We will follow a Renormalized Hamiltonian Approach (RHA) similar to 
that used for strongly interacting fermions\cite{zhang,das}
based on Gutzwiller approximation\cite{gutz}.
In this method, the strict constraint of no double occupancy
shows up in the renormalization factor for hopping. The hopping
(or band) renormalization factor is obtained by taking
the ratio of
hopping probability between two sites in the correlated 
space to that in the 
non-correlated space. The
hopping probability in the projected space (of no multiple
occupancy) is 
\begin{equation}
p_{corr}=n(1-n).
\end{equation}
The above equation means that the site from which hopping takes
place should be occupied and the target site should be empty
in the projected space. To find the  
hopping probability in the non-correlated state, we have to first
find the probability that a site is occupied. The occupancy
of the site to which a boson hops is irrelevant since we are dealing
with bosons.  To find the site occupancy probability of a site
from which a boson hops, we employ the following route.
First we find the number of ways $N_B$ number of bosons can be 
distributed in $N_S$ number of lattice sites and then
the number of ways of distributing the same number of bosons
in ($N_{S}-1$) number of lattice sites. 
The difference would give the number of configurations
where a particular site among $N_S$ number of lattice sites
is occupied by bosons.
The number of configurations possible for $N_{B}$ bosons on $N_S$ sites
is given by
\begin{equation}
w(N_{B}:N_{S})=\frac{(N_{B}+N_{S}-1)!}{N_{B}!(N_{S}-1)!}.
\end{equation}
The number of configurations possible for $N_{B}$ bosons on $(N_S-1)$ sites
is given by
\begin{equation}
w(N_{B}:N_{S}-1)=\frac{(N_{B}+N_{S}-2)!}{N_{B}!(N_{S}-2)!}.
\end{equation}
The probability that a particular site is occupied is then
\begin{equation}
p_{occ}=1-\frac{w(N_{B}:N_{S}-1)}{w(N_{B}:N_{S})}.
\end{equation}
In the thermodynamic limit,
\begin{equation}
p_{occ}=\frac{n}{1+n}.
\end{equation}
As mentioned for non-interacting bosons,
the hopping takes place irrespective of 
the target site being empty or occupied by any
number of bosons. 
Hence the hopping probability
for the non-interacting case is
\begin{equation}
p_{nocorr}=\frac{n}{1+n}.
\end{equation}
The band-width renormalization factor ($p_{corr}/p_{nocorr}$) then is,
\begin{equation}
\phi_{B}=1-n^{2}.
\end{equation}
Using the above $\phi_{B}$ the renormalized Hamiltonian valid in the strong coupling
limit and for $n\leq 1$ is,
\begin{equation}
H_{sc}=\sum_{{\bf k}}[\phi_{B}(n)\epsilon({\bf k})-\mu]
c^{\dag}_{{\bf k}}c_{{\bf k}}.
\end{equation}
The above is the Hamiltonian of Tonks gas on a lattice. One can immediately
see that the band width is strongly $n$ dependent. For $n=1$, the 
effective mass diverges and the system is a Bose-Mott-Hubbard insulator.
In passing , we note that for the fermion Hubbard model, the
corresponding hopping renormalization
factor\cite{zhang} is $2(1-n)/(2-n)$.
We do realize that the RHA approach based on Gutzwiller approximation
used to obtain Eq. 20 has its limitations. Within this approach,
the effect of strong interaction shows up only in the enhanced effective
mass of the bosons. This approximation would be reasonable
for large connectivity (i.e., for large number of nearest
neighbors). One can get some feel for the accuracy of the approximation
by comparing with the same for fermions. In the case fermions, it has been
shown that the Gutzwiller approximation gets progressively better
as the dimensionality increases and becomes exact in infinite dimensions and
that in three dimensions it is reasonable\cite{kot}. 
For strongly interacting bosons also, the Gutzwiller wave 
function is believed to be exact
for large dimensionality\cite{rok} and can be expected to be
reasonable in three dimensions. 
\par
As mentioned earlier the chemical potential and the Bose condensation
temperature ($T_B$) is determined by solving the number equation. Results of
this calculations, for bosons in a sc lattice,  are shown in Fig. 1.
For the non-interacting bosons, the $T_B$ shows an initial fast rise
and then almost a linear increase with $n$. In the case of strongly interacting
bosons, correlation-effects 
lead to an enhancement of the boson effective mass with
increasing band-filling. This increase in the
effective mass becomes dominant beyond $n=0.5$. As a consequence the $T_B$
starts decreasing and finally going to zero at integer filling for which
the transition to Bose-Mott-Hubbard insulating state takes place.
It may be noted that the $T_B$ for strongly interacting bosons
is almost the same as that for non-interacting bosons for
low filling ($n<0.2$). This is a signature of the fact that
correlation effects are not significant for low density.
The specific heat is calculated using the Eqs. (3) and (10).
For the Tonks gas calculations, the original band-width is replaced by the
renormalized band-width.
The results for non-interacting bosons in a sc lattice and a comparison 
with that for strongly interacting bosons are shown in Fig. 2.
In general, the specific heat increases with temperature for $T<T_B$,
shows a $\lambda$ anomaly at $T_B$ and then decreases with temperature.
For non-interacting bosons, the specific heat (per site) follows the
same curve at low temperatures irrespective of the band filling.
The peak value of the specific heat, however, depends on the filling.
The peak value is larger for higher filling. It is also found that 
the peak value does not depend on the bandwidth. 
For strongly interacting bosons Fig.2 shows that {\it for a fixed boson
density} (i) the peak in the specific heat occurs at a temperature 
lower than that for the non-interacting bosons, (ii) the peak value, 
however, remains the same for both interacting and non-interacting cases. 
The latter result is a consequence of the fact that the peak value of
the specific heat depends only on the boson density and not
on the boson effective mass (or bandwidth).
The first behavior is a combined effect of the reduced value of $T_B$ for 
interacting bosons compared to the non-interacting bosons at the same 
density and the appearance of the specific heat peak at $T_B$.   
As mentioned previously the effect of the on-site interation is small 
for low boson density and becomes increasingly dominant as the boson
density approaches to integer filling, the difference between the 
specific heat curves as well as between the peak temperatures for the 
interating and non-interacting bosons is small for $n=0.4$ and is large for
$n=0.8$ where the correlation effect is dominant. 
In experiments it is now possible to tune the hopping of bosons for a given
density of bosons to control the effective interaction strength and to
obtain results for weakly (or non-interacting) bosons and strongly
interacting bosons.  The ratio of the peak temperatures of the specific 
heat for the non-interacting and interacting cases for a fixed boson density 
may serve as a measure of effective interaction. 
A comparison of specific heats for non-interacting bosons in a sc
lattice and that for bosons in a 3-d
box  is shown in Fig. 3. The specific heat for free bosons in a 3-d
box is obtained by using the DOS $\rho(\epsilon)\propto \sqrt{\epsilon}$.
For the sc case the specific heat follows a S-shaped curve
for $T<T_B$ and it shows a considerable temperature 
dependence above $T_B$  compared to that of free bosons.
The $\lambda$ anomaly is very prominent for the sc case
in comparison with the free bosons.
The results for strongly interacting bosons are shown in Fig. 4.
For the strongly interacting bosons, it is
seen that with increasing $n$ the specific heat versus temperature 
curves become narrower. 
At a low temperature the specific heat 
as well as the slope of the specific heat with temperature increases 
with increasing $n$. This is because of increasing effective mass
of the bosons with $n$. 
\section{Conclusions}
\label{sec3}
In this paper we have presented specific heat for bosons
in a tight-binding band corresponding to a 
sc lattice. We investigated non-interacting bosons and Tonks gas.
In both cases, the specific heat is found to have considerably more
temperature dependence above $T_B$ compared to that of free
bosons in a box. For the strongly interacting bosons, we find that the
low-temperature specific heat gets enhanced as the systems 
moves towards the Mott transition.

\begin{figure}
\resizebox*{3.1in}{2.5in}{\rotatebox{270}{\includegraphics{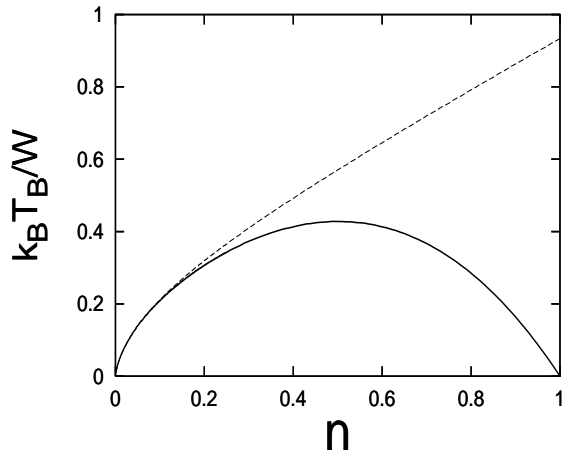}}}
\vspace*{0.5cm}
\caption[]{
Bose condensation temperature for non-interacting bosons (dots)
and strongly interacting bosons (solid line) as a function of number
density ($n$) for bosons in a sc lattice. 
Here $W$ is the half-band-width.
}
\label{scaling}
\end{figure}
\begin{figure}
\resizebox*{3.1in}{2.5in}{\rotatebox{270}{\includegraphics{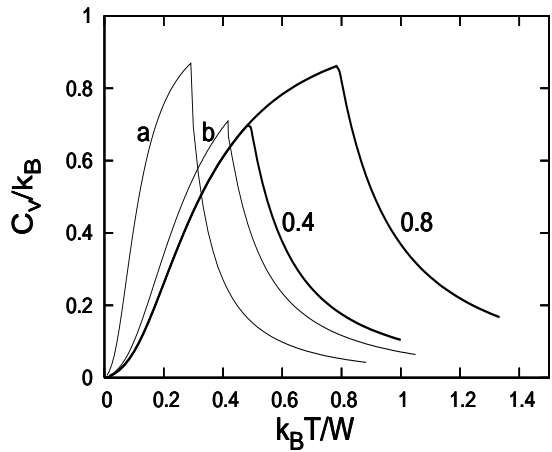}}}
\vspace*{0.5cm}
\caption[]{Temperature variation of specific heat per site for 
non-interacting bosons (thick lines) in a  sc lattice
for two values of $n$ ($0.4$ and $0.8$). Thin lines 
are for strongly interacting bosons, (a) $n = 0.8$ and (b) $n=0.4$.
}
\label{scaling}
\end{figure}
\begin{figure}
\resizebox*{3.1in}{2.5in}{\rotatebox{270}{\includegraphics{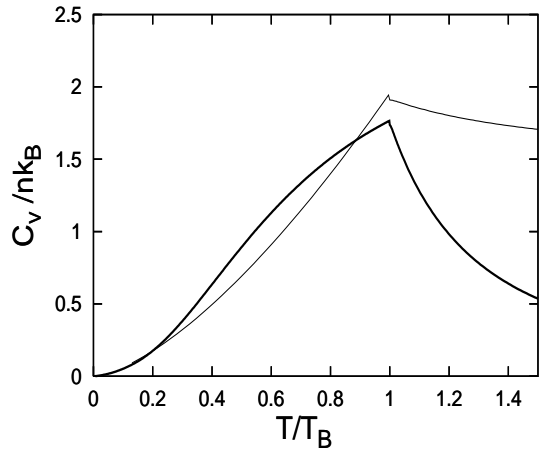}}}
\vspace*{0.5cm}
\caption[]{Temperature variation of specific heat
per particle for non-interacting bosons in  a sc lattice
(thick line) and for free bosons in a three dimensional box.}
\label{scaling}
\end{figure}
\begin{figure}
\resizebox*{3.1in}{2.5in}{\rotatebox{270}{\includegraphics{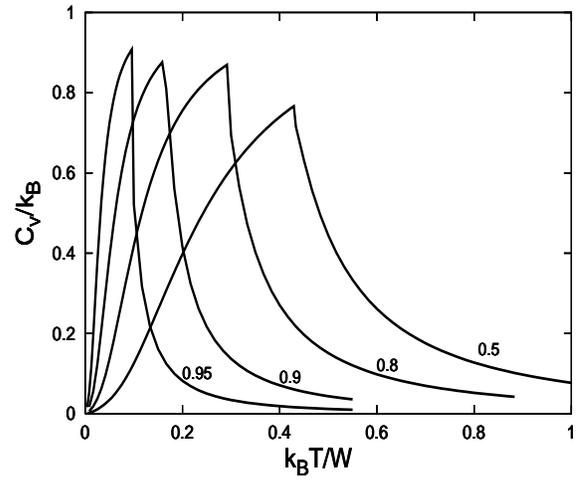}}}
\vspace*{0.5cm}
\caption[]{
Specific heat (per site) of Tonks gas in a sc lattice for several
values of $n$ shown on the curves. 
}
\label{scaling}
\end{figure}
\end{multicols}
\end{document}